\documentclass[prd,aps,amssymb,twocolumn,showpacs]{revtex4}
\usepackage{graphicx}
\usepackage{bm}
\usepackage{epsfig}
\usepackage{amsmath,amsfonts,mathrsfs,amssymb}
\usepackage{mathptmx}       
\usepackage{subfigure}
\usepackage{amssymb,amsfonts}
\usepackage{latexsym}
\newbox\pippobox

\begin{document}

\title{Enhancement of jet quenching around phase transition: \\
result from the dynamical holographic model}

\author{Danning Li$^{1}$,  Jinfeng Liao$^{2,3}$, Mei Huang $^{1,4}$ }
\affiliation{
$^{1}$ Institute of High Energy Physics, Chinese Academy of Sciences, Beijing, China \\
$^{2}$ Physics Department and Center for Exploration of Energy and Matter,
Indiana University, 2401 N Milo B. Sampson Lane, Bloomington, IN 47408, USA \\
$^{3}$ RIKEN BNL Research Center, Bldg. 510A, Brookhaven National Laboratory, Upton, NY 11973, USA \\
$^{4}$ Theoretical Physics Center for Science Facilities, Chinese
Academy of Sciences, Beijing, China }

\date{\today}

\begin{abstract}

The phase transition and jet quenching parameter ${\hat q}$ have
been investigated in the framework of dynamical holographic QCD model. It
is found that both the trace anomaly and the ratio of the jet quenching
parameter over cubic temperature ${\hat q}/T^3$ show a peak around the
critical temperature $T_c$, and the ratio of jet quenching
parameter over entropy density ${\hat q}/s$ sharply rises at $T_c$.
This indicates that the jet quenching parameter
can characterize the phase transition. The effect of jet quenching parameter
enhancement around phase transition on nuclear modification factor $R_{AA}$ and
elliptic flow $v_2$ have also been analyzed, and it is found that
the temperature dependent jet quenching parameter from dynamical holographic
QCD model can considerably improve the description of jet quenching azimuthal
anisotropy as compared with the conformal case.
\end{abstract}


\maketitle

\section{Introduction}

Studying Quantum Chromodynamics (QCD) phase transition and properties of
hot/dense quark matter at high temperature has been the main
target of heavy ion collision experiments at the Relativistic Heavy Ion
collider (RHIC) and the Large Hadron Collider (LHC). It is now believed
that the system created at RHIC and LHC is a strongly coupled quark-gluon
plasma (sQGP) and behaves like a nearly "perfect" fluid \cite{RHIC-EXP,RHIC-THEO}.

The collective flow $v_2$ of the highly excited and strongly interacting matter formed
at RHIC can be well described by relativistic hydrodynamics with a negligible ratio
of shear viscosity over entropy density $\eta/s$ ~\cite{Hydro}. Lattice QCD calculation
confirmed that $\eta/s$ for the purely gluonic plasma is rather small and in the range
of $0.1-0.2$ \cite{LAT-etas}. Shear viscosity $\eta$ characterizes how strongly particles
interact and move collectively in a many-body system. In general, the stronger the
interparticle interaction, the smaller the ratio of shear viscosity over entropy density.
Another unusual feature of the strongly interacting matter formed at RHIC is that the
emission of hadrons with large transverse momentum is strongly suppressed in central collisions~\cite{jet-quenching}. The suppression of hadrons at large transverse momentum
is normally referred to as jet quenching, which characterizes the squared average
transverse momentum exchange between the medium and the fast parton per unit path length~\cite{Baier:1996kr}. Current knowledge on jet quenching is that it is caused by
gluon radiation induced by multiple collisions of the leading parton with color charges
in the near-thermal medium \cite{Baier:1996kr,energy-loss,GuoW}. Therefore, jet quenching
can tell us the properties of the created hot dense matter by the energetic
parton passing through the medium.

It is very interesting to ask whether transport quantities can characterize phase transitions.

It has been observed that the shear viscosity over entropy density ratio $\eta/s$ has a
minimum in the phase transition region in systems of water, helium, nitrogen \cite{Csernai:2006zz} and many other substances \cite{Liao:2009gb}. It has been shown
$\eta /s$ is suppressed near the critical temperature in the semi quark gluon plasma \cite{Hidaka:2009ma}, and $\eta /s$ can characterize first-, second-order phase transitions and crossover \cite{etas-scalar}, i.e. $\eta /s$ shows a cusp, a jump at $T_c$ and a
shallow valley around $T_c$, respectively.

In \cite{Majumder:2007zh} it has been suggested that the jet quenching parameter can also be
used to measure the coupling strength of the medium and a general relation between the shear viscosity $\eta/s$ and the jet quenching parameter $\hat{q}/T^3$ for a quasi-particle dominated quark-gluon plasma has been derived, i.e. $\eta/s \sim T^3/\hat{q}$.
The relation associates a small ratio of shear viscosity to entropy density to a
large value of the jet quenching parameter. If we naively extend this relation
to the critical temperature region, we would expect that $\hat{q}/T^3$ will show a peak around
the critical temperature $T_c$. Phenomenologically, the strong near-Tc-enhancement (NTcE) scenario of jet-medium interaction ~\cite{Liao:2008dk} was proposed in the efforts
to explain the large jet quenching anisotropy at high $p_t$ at RHIC~\cite{star2005,phenix2010,Jia:2010ee,Jia:2011pi}. More recently it was shown in Refs.~\cite{Zhang:2012ie,Zhang:2012ha} that the NTcE model naturally induces a reduction
($\sim 30\%$) of jet-medium interaction strength from RHIC to LHC.

It is worthy of mentioning that another transport coefficient, the bulk viscosity $\zeta/s$,
also exhibits a sharp rising behavior around the critical temperature $T_c$ as shown in
Lattice QCD \cite{LAT-xis-KT,LAT-xis-Meyer,correlation-Karsch}, the linear
sigma model \cite{bulk-Paech-Pratt}, the Polyakov-loop linear sigma model \cite{Mao:2009aq}
and the real scalar model \cite{Li-Huang}.
The rising of bulk viscosity near phase transition corresponds to peak of trace anomaly
around $T_c$, which shows the equation of state is highly non-conformal \cite{LAT-EOS-G, LAT-EOS-Nf2} around phase transition.

Due to the complexity of QCD in the regime of strong coupling, in recent years, the
anti-de Sitter/conformal field theory (AdS/CFT) correspondence \cite{Maldacena:1997re,Gubser:1998bc,Witten:1998qj}
has generated enormous interest in using thermal ${\cal N} = 4$ super-Yang-Mills theory
(SYM) to understand sQGP. However, a conspicuous
shortcoming of this approach is the conformality of SYM: the square of the speed of
sound $c_s^2$ always equals to $1/3$, the bulk viscosity is always zero at all
temperatures in this theory, and $\eta/s=\frac{1}{4\pi}$ \cite{bound}
and $\hat{q}/T^3\simeq 7.53 \sqrt{\lambda}$  ($\lambda=g_{YM}^2N_c$ the 't Hooft
coupling) \cite{Liu:2006ug} keeps a constant for all temperatures. In order to
describe the nonconformal properties near phase transition, and mimic the QCD
equation of state, much effort has been put to find the gravity dual of gauge
theories which break the conformal symmetry, e.g,
\cite{Gubser-T,Gursoy-T,Pirner-T,Li:2011hp}, where a real scalar dilaton field
background has been introduced to couple with the graviton.
Refs.\cite{Gubser-T} and \cite{Gursoy-T} have used different dilaton
potentials as input, Ref.\cite{Pirner-T} has used QCD
$\beta$-function as input, and Ref.\cite{Li:2011hp} has introduced
a deformed metric background.

On the other hand, much efforts have also been paid to establish a more
realistic holographic QCD model for glueball spectra and meson spectra
\cite{EKSS2005,Karch:2006pv,Csaki:2006ji,Gherghetta-Kapusta-Kelley,YLWu}.
Recently, a dynamical holographic QCD model \cite{Li:2012ay,Li:2013oda,Li:2013xpa}
has been developed by resembling the renormalization group from ultraviolet (UV)
to infrared (IR). The dynamical holographic QCD model is constructed in the
graviton-dilaton-scalar
framework, where the dilaton background field $\Phi$ and scalar field $X$ are
responsible for the gluodynamics and chiral dynamics, respectively. At the UV boundary,
the dilaton field is dual to the dimension-4 gluon operator, and the scalar field is
dual to the dimension-3 quark-antiquark operator. The metric structure at IR is
automatically deformed by the nonperturbative gluon condensation and chiral
condensation in the vacuum. The produced scalar glueball spectra in the
graviton-dilaton framework agree well with lattice data, and the light-flavor
meson spectra generated in the graviton-dilaton-scalar framework are in well
agreement with experimental data. Both the chiral symmetry breaking and linear
confinement are realized in this dynamical holographic QCD model.

In this work, we will investigate the phase transition, equation of state
and calculate the jet quenching parameter in the dynamical holographic QCD model.
The paper is organized as follows. In Sec.\ref{sec-model}, we briefly introduce
the dynamical holographic QCD model for pure gluon system and light-flavor system. In
Sec.\ref{sec-EOS}, we will investigate the phase transition and equation of state,
including the entropy density, the pressure density, the energy density
for the pure gluon system. Then in Sec.\ref{sec-jet}, Sec.\ref{sec-q-hat} and Sec.\ref{sec-RAA-v2},
we calculate jet quenching parameter and investigate the nuclear modification factor
$R_{AA}$ and elliptic flow $v_2$. The summary and discussion is
given in Sec.\ref{sec-summary}.

\section{Dynamical holographic QCD model}
\label{sec-model}

Quantum chromodynamics (QCD) in terms of quark and gluon degrees of freedom
is accepted as the fundamental theory of the strong interaction. In the
ultraviolet (UV) or weak coupling regime of QCD, the perturbative calculations
for deep inelastic scattering (DIS) agree well with experimental data. However,
in the infrared (IR) regime, the description of QCD vacuum as well as hadron
properties and nonperturbative processes still remains as outstanding challenge
in the formulation of QCD as a local quantum field theory.
In the past half century, various non-perturbative methods have been developed, in
particular lattice QCD, Dyson-Schwinger equations (DSEs), and functional
renormalization group equations (FRGs). In recent decades, an entirely new method
based on the anti-de Sitter/conformal field theory (AdS/CFT) correspondence and the
conjecture of the gravity/gauge duality \cite{Maldacena:1997re,Gubser:1998bc,Witten:1998qj}
provides a revolutionary method to tackle the problem of strongly coupled gauge theories.

In general, holography relates quantum field theory (QFT) in d-dimensions to quantum
gravity in (d + 1)-dimensions, with the gravitational description becoming classical
when the QFT is strongly-coupled. The extra dimension can be interpreted as an energy
scale or renormalization group (RG) flow in the QFT \cite{Adams:2012th}.

\begin{figure}[!htb]
\begin{center}
\includegraphics[scale=.4]{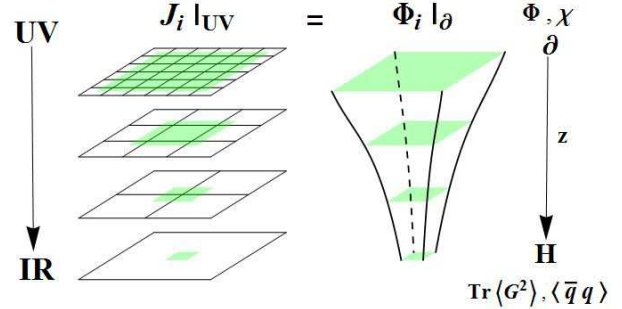}
\caption{Duality between $d$-dimension QFT and $d+1$-dimension gravity as shown
in \cite{Adams:2012th} (Left-handed side). Dynamical holographic QCD model resembles
RG from UV to IR (Right-handed side): at UV boundary the dilaton bulk field $\Phi(z)$
and scalar field $X(z)$ are dual to the dimension-4 gluon operator and dimension-3
quark-antiquark operator, which develop condensates at IR. }
\label{fig:RGflow}
\end{center}
\end{figure}

The recently developed dynamical holographic QCD model \cite{Li:2012ay,Li:2013oda}
can resemble the renormalization group from ultraviolet (UV) to infrared (IR) as
shown in Fig.\ref{fig:RGflow} \cite{Li:2013xpa}. The dilaton background $\Phi(z)$
is introduced to describe the gluodynamics, and the scalar field $X(z)$ is responsible
for chiral dynamics, respectively.

For the pure gluon system, we construct the quenched dynamical holographic QCD model
in the graviton-dilaton framework by introducing one scalar dilaton
field $\Phi(z)$ in the bulk. The 5D graviton-dilaton coupled action in the string frame
is given below:
\begin{eqnarray}
 S_G=\frac{1}{16\pi G_5}\int
 d^5x\sqrt{g_s}e^{-2\Phi}\left(R_s+4\partial_M\Phi\partial^M\Phi-V^s_G(\Phi)\right).
\label{action-gluon}
\end{eqnarray}
Where $G_5$ is the 5D Newton constant, $g_s$, $\Phi$ and $V_G^s$ are the 5D
metric, the dilaton field and dilaton potential in the string frame, respectively.
The metric ansatz is often chosen to be
\begin{eqnarray}\label{metric-ansatz}
ds^2=e^{2A_s(z)}(dz^2+\eta_{\mu\nu}dx^\mu dx^\nu).
\end{eqnarray}
In this paper, the capital letters like "M,N" would stand for all the coordinates(0,1,..,4), and the greek indexes would stand for the 4D coordinates(0,...,3). We would use the convention $\eta^{00}=\eta_{00}=-1,\eta^{ij}=\eta_{ij}=\delta_{ij}$.)

To avoid the gauge non-invariant problem and to meet the requirement of gauge/gravity
duality, we take the dilaton field in the form of
\begin{equation}
\Phi(z)=\mu_G^2z^2\tanh(\mu_{G^2}^4z^2/\mu_G^2).
\label{mixed-dilaton}
\end{equation}
In this way, the dilaton field at UV behaves
$\Phi(z)\overset{z\rightarrow0}{\rightarrow} \mu_{G^2}^4 z^4$,
and is dual to the dimension-4 gauge invariant gluon operator ${\rm Tr} G^2 $,
while at IR it takes the quadratic form
$\Phi(z)\overset{z\rightarrow\infty}{\rightarrow} \mu_G^2 z^2$.
The equations of motion can be derived as
\begin{eqnarray}
-A_s^{''}-\frac{4}{3}\Phi^{'}A_s^{'}+A_s^2+\frac{2}{3}\Phi^{''}=0, \\
\Phi^{''}+(3A_s^{'}-2\Phi^{'})\Phi^{'}-\frac{3}{8}e^{2A_s-\frac{4}{3}\Phi}\partial_\Phi (e^{\frac{4}{3}\Phi}V_G^s(\Phi))=0.
\end{eqnarray}
By self-consistently
solving the Einstein equations, the metric structure $A_s$ will be automatically
deformed at IR by the dilaton background field or the nonperturbative gluodynamics.
It is found in \cite{Li:2013oda} that the scalar glueball spectra in the
quenched dynamical model is in very well agreement with lattice data. For
details, please refer to Refs.\cite{Li:2013oda}.

To describe the two-flavor system, we then add light flavors in terms of meson fields
on the gluodynamical background. The total 5D action for the graviton-dilaton-scalar
system takes the following form:
\begin{eqnarray}
 S=S_G + \frac{N_f}{N_c} S_{KKSS}.
\end{eqnarray}
Here $S_G$ is the 5D action for gluons in terms of dilaton field $\Phi$ and
takes the same form as Eq.(\ref{action-gluon}), and $S_{KKSS}$ is the 5D action
for mesons propagating on the dilaton background and takes the same form as
in the KKSS model \cite{Karch:2006pv}
\begin{eqnarray}
S_{KKSS}&=&-\int d^5x
 \sqrt{g_s}e^{-\Phi}Tr(|DX|^2+V_X(X^+X, \Phi) \nonumber \\
 && ~~+\frac{1}{4g_5^2}(F_L^2+F_R^2)).
\end{eqnarray}
The difference here is that the metric structure $A_s$ is solved
from the following coupled equations of motion:
\begin{eqnarray}
 -A_s^{''}+A_s^{'2}+\frac{2}{3}\Phi^{''}-\frac{4}{3}A_s^{'}\Phi^{'}
 -\frac{\lambda_0}{6}e^{\Phi}\chi^{'2}&=&0, \label{Eq-As-Phi} \\
 \Phi^{''}+(3A_s^{'}-2\Phi^{'})\Phi^{'}-\frac{3\lambda_0}{16}e^{\Phi}\chi^{'2} & &
 \nonumber \\
 -\frac{3}{8}e^{2A_s-\frac{4}{3}\Phi}\partial_{\Phi}\left(V_G(\Phi)
 +\lambda_0 e^{\frac{7}{3}\Phi}V_C(\chi,\Phi)\right)&=&0, \label{Eq-VG}\\
 \chi^{''}+(3A_s^{'}-\Phi^{'})\chi^{'}-e^{2A_s}V_{C,\chi}(\chi,\Phi)&=&0. \label{Eq-Vc}
\end{eqnarray}
Here we have defined $V_C=  Tr(V_X)$ and $V_{C,\chi}=\frac{\partial V_C}{\partial \chi}$.
$ \frac{16\pi G_5 N_f}{L^3 N_c}\rightarrow \lambda_0 $.

For two flavor system in the graviton-dilaton-scalar framework, the
deformed metric is self-consistently solved by considering both the chiral condensate
and nonperturbative gluodynamics in the vacuum, which are responsible for the chiral
symmetry breaking and linear confinement, respectively. The mixing
between the chiral condensate and gluon condensate is important to produce the correct
light flavor meson spectra \cite{Li:2013oda}.

\section{Phase transition and equation of states}
\label{sec-EOS}

The chiral and deconfinement phase transitions for the graviton-dilaton-scalar system
will be investigated in the near future.
In this section, we will focus on the deconfinement phase transition
for the pure gluon system described by Eq.(\ref{action-gluon}).

The finite temperature dynamics of gauge theories, has a natural
holographic counterpart in the thermodynamics of black-holes on the
gravity side. Adding the black-hole background to the holographic QCD model
constructed from vacuum properties, in the string frame we have
\begin{equation} \label{metric-stringframe}
ds_S^2=
e^{2A_s}\left(-f(z)dt^2+\frac{dz^2}{f(z)}+dx^{i}dx^{i}\right).
\end{equation}
The metric in the string frame is useful to calculate the jet quenching
parameter.

The thermodynamical properties of equation of state is convenient
to be derived in the Einstein frame,
\begin{eqnarray} \label{metric-Einsteinframe}
ds_E^2= e^{2A_s-\frac{4\Phi}{3}}\left(-f(z)dt^2
+\frac{dz^2}{f(z)}+dx^{i}dx^{i}\right).
\label{Einstein-metric}
\end{eqnarray}
Under the frame transformation
\begin{equation}
g^E_{mn}=g^s_{mn}e^{-2\Phi/3}, ~~ V^E_G=e^{4\Phi/3}V_{G}^s,
\end{equation}
Eq.(\ref{action-gluon}) becomes
\begin{eqnarray}\label{graviton-dilaton-E}
S_G^E=\frac{1}{16\pi G_5}\int d^5x\sqrt{g_E}\left(R_E-\frac{4}{3}\partial_m\Phi\partial^m\Phi-V_G^E(\Phi)\right).
\end{eqnarray}

We can derive the following equations from the Einstein equations of $(t,t), (z,z)$ and $(x_1, x_1)$
components:
\begin{eqnarray}
 -A_s^{''}+A_s^{'2}+\frac{2}{3}\Phi^{''}-\frac{4}{3}A_s^{'}\Phi^{'}&=&0, \label{Eq-As-Phi-T} \\
 f''(z)+\left(3 A_s'(z) -2 \Phi '(z)\right)f'(z)&=&0\label{Eq-As-f-T}
\end{eqnarray}

The EOM of the dilaton field is given as following
\begin{equation}
\label{fundilaton} \frac{8}{3} \partial_z
\left(e^{3A_s(z)-2\Phi} f(z)
\partial_z \Phi\right)-
e^{5A_s(z)-\frac{10}{3}\Phi}\partial_\Phi V_G^E=0.
\end{equation}

To get the solutions we impose the asymptotic
$AdS_5$ condition $f(0)=1$ near the UV boundary $z\sim 0$,  and
require $\Phi, f$ to be finite at $z=0, z_h$ with $z_h$ the black-hole
horizon. Fortunately, we find that the solution of the black-hole
background takes the form of,
\begin{eqnarray} \label{solu-f}
f(z)= 1- f_{c}^h \int_0^{z} e^{-3A_s(z^{\prime})+2\Phi(z^{\prime})} dz^{\prime},
\end{eqnarray}
with
\begin{eqnarray}\label{fc}
f_{c}^h= \frac{1}{\int_0^{z_h} e^{-3A_s(z^{\prime})+2\Phi(z^{\prime})} dz^{\prime} },
\end{eqnarray}.

A black-hole solution with a regular horizon is characterized by the
existence of a surface $z=z_h$, where $f(z_h)=0$. The Euclidean
version of the solution is defined only for $0 < z < z_h$, in order
to avoid the conical singularity, the periodicity of the Euclidean
time can be fixed by
\begin{equation}
\tau \rightarrow \tau+\frac{4\pi}{|f'(z_h)|}.
\end{equation}
This determines the temperature of the solution as
\begin{equation}
T=\frac{|f'(z_h)|}{4\pi}.
\end{equation}

From Eq. (\ref{solu-f}), one can easily read out the relation
between the temperature and position of the black hole horizon.
\begin{equation} \label{temp}
T =\frac{e^{-3A_s(z_h)+2\Phi(z_h)}}{4\pi \int_0^{z_h} e^{-3A_s(z^{\prime})+2\Phi(z^{\prime})} dz^{\prime} }
\end{equation}

For numerical calculations, we take $\mu_G=0.75 {\rm GeV}$ in Eq.(\ref{mixed-dilaton})
so that the transition temperature is around $255 {\rm MeV}$, and we take three different
values for $\mu_{G^2}$: $\mu_{G^2}=\mu_G=0.75 {\rm GeV}$, $\mu_{G^2}=3 {\rm GeV}$ and $\mu_{G^2}=\infty$. When $\mu_{G^2}=\infty$, the dilaton field Eq.(\ref{mixed-dilaton})
takes the form of quadratic term $\Phi=\mu_G^2 z^2$, and the model can be regarded as the self-consistent KKSS model. The only difference is that in this model the metric structure
is selfconsistently deformed by the dilaton background, while in the KKSS model, the metric
structure takes the same as $AdS_5$.

\begin{figure}[h]
\begin{center}
\epsfxsize=6.5 cm \epsfysize=6.5 cm \epsfbox{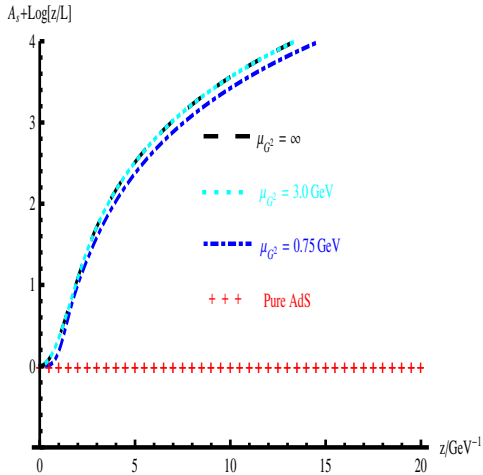}
\end{center}
\caption{$A_s$ configurations compared with $AdS_5$ metric for $G_5=1.25$ and
$\mu_G=0.75 {\rm GeV}$ and $\mu_{G^2}=\mu_G=0.75 {\rm GeV}$, $\mu_{G^2}=3 {\rm GeV}$ and $\mu_{G^2}=\infty$, respectively. To show the configuration smoothly, we have subtracted the $\log(z)$ divergence in $A_s$.} \label{Fig-As}
\end{figure}

We can solve $A_s$ from Eq.(\ref{Eq-As-Phi-T}), and the results of $A_s$ configurations
for $\mu_G=0.75 {\rm GeV}$ and $\mu_{G^2}=0.75 {\rm GeV}$, $\mu_{G^2}=3 {\rm GeV}$ and $\mu_{G^2}=\infty$ are shown in Fig.\ref{Fig-As}. To show the configuration smoothly, we have subtracted the $\log(z)$ divergence in $A_s$. Comparing with $AdS_5$ metric, it is easy to find that the metric structure is largely deformed at IR by the dilaton background field or gluodynamics. The two cases $\mu_{G^2}=3 {\rm GeV}$ and $\mu_{G^2}=\infty$ are
almost the same.

\begin{figure}[h]
\begin{center}
\epsfxsize=6.5 cm \epsfysize=6.5 cm \epsfbox{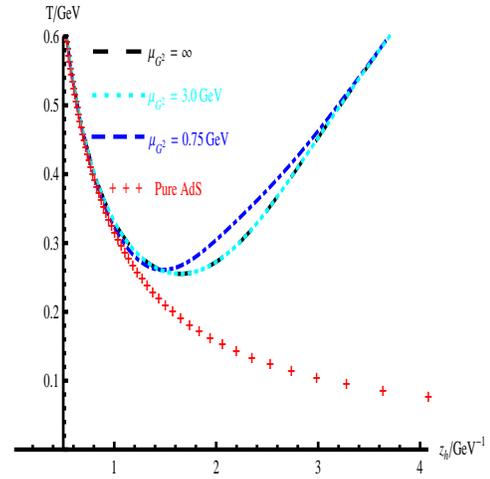}
\end{center}
\caption{The temperature as a function of horizon for $G_5=1.25$ and $\mu_G=0.75 {\rm GeV}$ and $\mu_{G^2}=0.75 {\rm GeV}$, $\mu_{G^2}=3 {\rm GeV}$ and $\mu_{G^2}=\infty$,
respectively. The blue lines stand for $Case~I,II$
and the red lines are results of AdS-Schwardz black hole.} \label{Tzh}
\end{figure}

From Eq.(\ref{Eq-As-f-T}) and Eq.(\ref{temp}), we can get the
$f(z)$ solution and the temperature behavior.
The temperature v.s. horizon for $\mu_G=0.75 {\rm GeV}$ and $\mu_{G^2}=0.75 {\rm GeV}$, $\mu_{G^2}=3 {\rm GeV}$ and $\mu_{G^2}=\infty$ are shown in Fig.\ref{Tzh}. As long as $\mu_{G^2}$ is large, the IR physics is not sensitive to large $\mu_{G^2}$, the
behaviors for $\mu_{G^2}=3 {\rm GeV}$ and $\mu_{G^2}=\infty$ are almost the same.
From Fig. \ref{Tzh}, it is noticed that for pure ${\rm AdS}_5$
Schwarz black-hole, the temperature monotonically decreases with the
increasing of the horizon. If
one solves the dual black-hole background self-consistently, one can
find that there is a minimal temperature $T_{min}=255 {\rm MeV}$ at certain
black-hole horizon $z_h^0$. This is similar to the case for the
confining theory (at zero temperature) discussed in Ref.
\cite{Gursoy-T}. For $T<T_{min}$, there are no black-hole
solutions. For $T>T_{min}$, there are two branches of black-hole
solutions. When $z_h<z_h^0$, the temperature increases with the
decreasing of $z_h$, which means that the temperature increases when
the horizon moves close to UV, this phase is thermodynamically
stable. When $z_h>z_h^0$, the temperature increases with the
increase of $z_h$, which means that the temperature becomes higher
and higher when the horizon moves to IR. This indicates that the
solution for the branch $z_h>z_h^0$ is unstable and thus not
physical. In order to determine the critical temperature, we have to compare
the free energy difference between the stable black hole solution
and the thermal gas. Following the discussion in \cite{Li:2011hp},
the transition temperature would be near this minimal temperature and
we would just take it as the transition temperature $T_c=255 {\rm MeV}$,
which is in agreement with lattice result for pure gluon system.

\begin{figure}[h]
\begin{center}
\epsfxsize=6.5 cm \epsfysize=6.5 cm \epsfbox{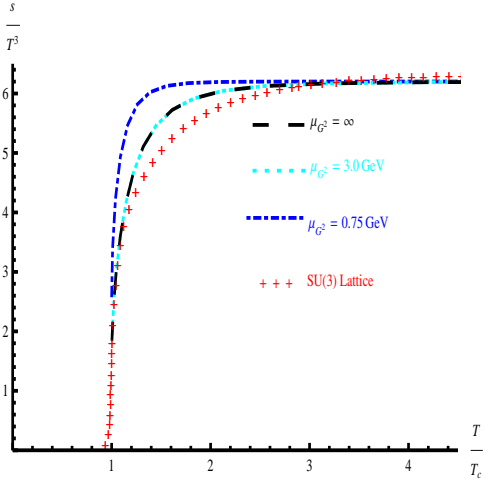}
\end{center}
\caption{The entropy density as a function of $T/T_c$ for $G_5=1.25$ and
$\mu_{G}=0.75 {\rm GeV}$ and $\mu_{G^2}=0.75 {\rm GeV}$, $\mu_{G^2}=3 {\rm GeV}$ and $\mu_{G^2}=\infty$,
respectively. The red crosses are lattice results from \cite{LAT-EOS-G}. } \label{SdT3}
\end{figure}

From the Bekenstein-Hawking formula, one can
easily read the black-hole entropy density $s$, which is defined by
the area $A_{area}$ of the horizon:
\begin{eqnarray}
s=\frac{Area}{4G_5V_3}|_{z_h}=\frac{1}{4G_5}e^{3A_s(z_h)-2\Phi(z_h)}.
\end{eqnarray}
Where $G_5$ is the Newton constant in 5D curved space and $V_3$ is
the volume of the spatial directions. It is noticed that the entropy
density is closely related to the metric in the Einstein frame.
The results of scaling entropy density $s/T^3$ for $\mu_{G}=0.75 {\rm GeV}$ and
$\mu_{G^2}=0.75 {\rm GeV}$, $\mu_{G^2}=3 {\rm GeV}$ and $\mu_{G^2}=\infty$
are shown in Fig.\ref{SdT3} compared with lattice results for the pure gluon system \cite{LAT-EOS-G}. It can be seen that when $\mu_{G^2}$ is large enough, the result
is not sensitive to the value of $\mu_{G^2}$, and it takes almost the same as that in
the selfconsist KKSS model. The entropy density for $\mu_{G^2}= 3 {\rm GeV}$ to $\mu_{G^2}=\infty$ agrees well with the lattice result for pure $SU(3)$ gauge theory.

The pressure density $p(T)$ can be calculated from the entropy
density $s(T)$ by solving the equation
\begin{equation}
\frac{dp(T)}{dT}= s(T),
\end{equation}
and the energy density is related to the entropy density by
\begin{eqnarray}
\epsilon=-p+sT.
\end{eqnarray}

\begin{figure}[h]
\begin{center}
\epsfxsize=6.5 cm \epsfysize=6.5 cm \epsfbox{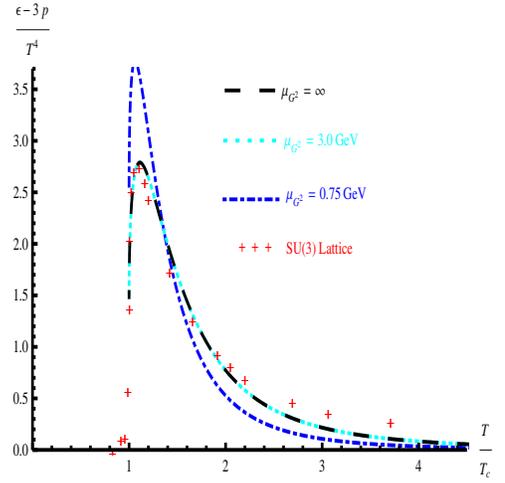}
\end{center}
\caption{Trace anomaly as a function of $T/T_c$ for $G_5=1.25$ and $\mu_{G}=0.75 {\rm GeV}$ and $\mu_{G^2}=0.75 {\rm GeV}$, $\mu_{G^2}=3 {\rm GeV}$ and $\mu_{G^2}=\infty$,
respectively. The red crosses are lattice results from \cite{LAT-EOS-G}.}
\label{TraceAnomaly}
\end{figure}

The trace anomaly $(\epsilon-3p)/T^4$ for $\mu_{G}=0.75 {\rm GeV}$ and
$\mu_{G^2}=0.75 {\rm GeV}$, $\mu_{G^2}=3 {\rm GeV}$ and $\mu_{G^2}=\infty$
are shown in Fig.\ref{TraceAnomaly} compared with lattice results for the
pure gluon system \cite{LAT-EOS-G}. The trace anomaly shows a peak around
$T/T_{c}=1.1$. When $\mu_{G^2}=0.75 {\rm GeV}$, the height of the peak
is around $3.7$, and when $\mu_{G^2}=3 {\rm GeV} \sim \infty$, the height
reduces to $2.7$, which agrees with lattice data for pure $SU(3)$ gauge theory
as shown in \cite{LAT-EOS-G}. At very high temperature, the trace anomaly goes
to zero, which indicates the system is asymptotically conformal at high temperature.

The sound velocity $c_s^2$ can be obtained from the temperature and
entropy:
\begin{equation}\label{sound}
c_s^2=\frac{d \log T}{d \log s}=\frac{s}{T ds/dT},
\end{equation}
which can directly measure the conformality of the system. For
conformal system, $c_s^2=1/3$, for non-conformal system, $c_s^2$
will deviate from $1/3$. From Eq.(\ref{sound}), we can see that the
speed of the sound is independent of the normalization of the 5D
Newton constant $G_5$ and the space volume $V_3$.

\begin{figure}[h]
\begin{center}
\epsfxsize=6.5 cm \epsfysize=6.5 cm \epsfbox{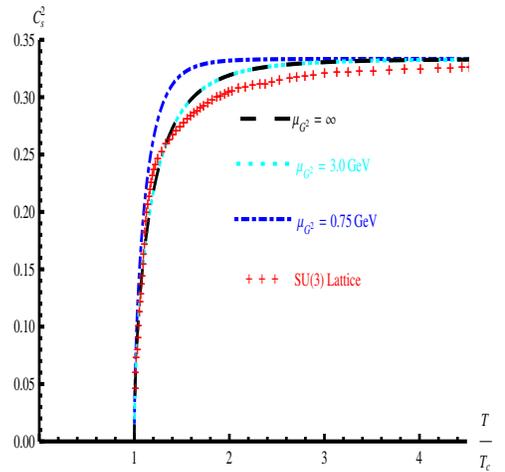}
\end{center}
\caption{The square of the sound velocity $c_s^2$ as a function of
scaled temperature $T/T_c$ for $G_5=1.25$ and $\mu_{G}=0.75 {\rm GeV}$
and $\mu_{G^2}=0.75 {\rm GeV}$, $\mu_{G^2}=3 {\rm GeV}$ and $\mu_{G^2}=\infty$,
respectively. The red crosses are lattice results from \cite{LAT-EOS-G}.} \label{SoundSpeed}
\end{figure}

The numerical result of the square of the sound velocity is shown in
Fig.\ref{SoundSpeed}. At $T_{c}$, the sound velocity square is around
$0$ which is in agreement with lattice data \cite{LAT-EOS-G}.
At high temperature, the sound velocity square
goes to $1/3$, which means that the system is asymptotically
conformal.

\section{Jet quenching parameter ${\hat q}$}
\label{sec-jet}

Jet quenching measures an energetic parton interacts with the created hot dense medium.
It is very important to find the characterization of the resulting medium induced modification of high-$p_T$ parton fragmentation, i.e., jet quenching and
its connection to properties of the hot dense matter, and whether and how
such a parameter can tell us about the QCD phase transitions.

It has been expected that the shear viscosity over entropy density ratio $\eta/s$ has
a minimum in the QCD phase transition region \cite{etas-scalar} as that in systems of water, helium and nitrogen \cite{Csernai:2006zz,Liao:2009gb}. The bulk viscosity $\zeta/s$,
also exhibits a sharp rising behavior around the critical temperature $T_c$ as shown in
Lattice QCD \cite{LAT-xis-KT,LAT-xis-Meyer,correlation-Karsch} and some model
calculations \cite{bulk-Paech-Pratt,Li-Huang}. It has been suggested in \cite{Majumder:2007zh} shear viscosity $\eta/s$ and the jet quenching parameter $\hat{q}/T^3$ for a quasi-particle dominated quark-gluon plasma has a general relation $\eta/s \sim T^3/\hat{q}$. If we naively extend this relation
to the critical temperature region, we would expect that $\hat{q}/T^3$ will show a
peak around the critical temperature $T_c$. Phenomenologically, the strong
near-Tc-enhancement (NTcE) scenario of jet-medium interaction ~\cite{Liao:2008dk}
was proposed in the efforts to explain the large jet quenching anisotropy at high
$p_t$ at RHIC~\cite{star2005,phenix2010,Jia:2010ee,Jia:2011pi}.

There has no model calculations for the jet quenching parameter around the critical
temperature. Lattice QCD is not suitable for transport properties,
recently there are several groups played efforts on calculate the jet quenching
parameter on lattice \cite{Majumder:2012sh,Panero:2013pla}. However, no information
on jet quenching parameter has been extracted around the critical temperature.
In this section, we will investigate the jet quenching parameter around the critical
temperature in the dynamical holographic QCD model which can describe phase transitions.

Following \cite{Liu:2006ug}(see also \cite{Gursoy:2009kk,Cai:2012eh,Zhang:2012jd}), the jet quenching parameter is related to the
adjoint Wilson loop by
\begin{eqnarray}
W^{Adj}[\mathcal {C}]\approx exp(-\frac{1}{4\sqrt{2}}\hat{q}L^{-}L^2)
\end{eqnarray}
where $L^{-},L$ are distances along $x^{-}=\frac{t-x_1}{\sqrt{2}}$ and
spatial direction $x_2$ respectively. (Another method for
jet quenching of light quarks has been developed in \cite{Ficnar:2013qxa}.)

Denoting $x^{+}=\frac{t+x_1}{\sqrt{2}},x^{-}=\frac{t-x_1}{\sqrt{2}}$,
then the metric in Eq.(\ref{metric-stringframe}) becomes
\begin{eqnarray}\label{metric-trans}
ds^2 &=& e^{2A_s}\{[1+f(z)]dx^{-}dx^{+}+\frac{1-f(z)}{2} \nonumber \\
     & & [dx^{+2}+dx^{-2}]+\frac{1}{f(z)}dz^2+dx_2^2+dx_3^2\},
\end{eqnarray}.

The action on the string world sheet is taken to be
\begin{eqnarray}
S_{NG}=\frac{1}{2\pi \alpha^{'}}\int d^2\sigma\sqrt{-det(G_{\alpha\beta})}
\end{eqnarray}
with $G_{\alpha \beta}=g^s_{\mu\nu}\partial_\alpha x^\mu\partial_\beta x^\nu$ the
induced metric on the string world sheet.

\begin{figure}[h]
\begin{center}
\epsfxsize=6.5 cm \epsfysize=6.5 cm \epsfbox{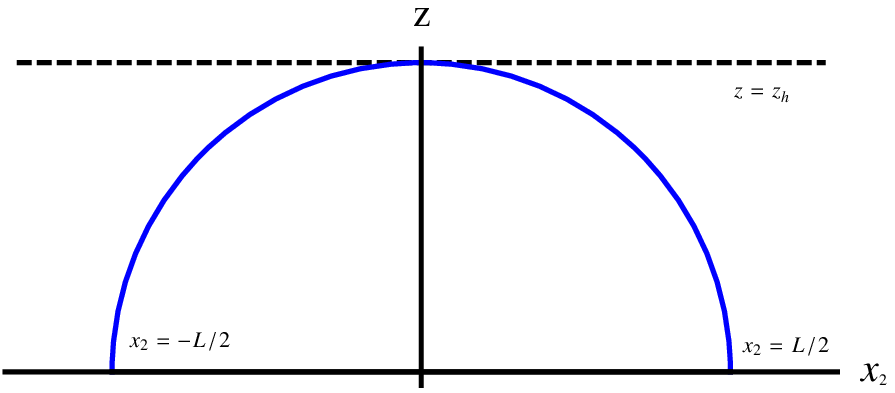} \\
\hskip 0.15 cm \textbf{( $\tau=x^{-},\sigma=x_2$ ) } \\
\hspace*{0.2cm}
\epsfxsize=6.5 cm \epsfysize=6.5 cm \epsfbox{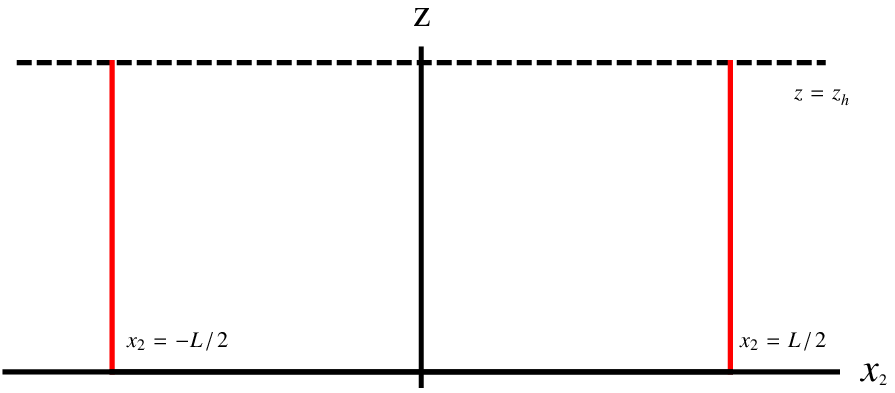} \\
\hskip 0.15 cm  \textbf{( $\tau=x^{-},\sigma=z$ )}
\end{center}
\caption{Two kinds of string configurations.} \label{StringConfiguration}
\end{figure}

With the above configuration in Fig.{\ref{StringConfiguration}} we have
\begin{eqnarray}
S_1=\frac{1}{2\pi\alpha^{'}}\int d\tau d\sigma\sqrt{g_{--}g_{zz}z^{'2}(\sigma)+g_{--}g_{22}} ,
\end{eqnarray}
and with the below configuration in Fig.{\ref{StringConfiguration}} we have
\begin{eqnarray}
S_2=\frac{1}{2\pi\alpha^{'}}\int d\tau d\sigma\sqrt{g_{--}g_{zz}},
\end{eqnarray}
where $g_{--}=e^{2A_s}\frac{1-f(z)}{2}, g_{zz}=\frac{e^{2A_s}}{f(z)}, g_{22}=e^{2A_s}$ can be read from Eq.(\ref{metric-trans}).

Then we extract the adjoint Wilson loop by
\begin{eqnarray}
W^{Adj}={\rm exp}(2i(S_1-S_2)),
\end{eqnarray}
and from the small $L$ expansion of $W^{Adj}$ we get $\hat{q}$ of the form
\begin{eqnarray}
\hat{q}=\frac{\sqrt{2}\sqrt{\lambda}}{\pi\int_0^{z_h}dz\sqrt{g_{zz}/(g^2_{22}g_{--})}},
\end{eqnarray}
with $\sqrt{\lambda}=\frac{R_{ads}^2}{\alpha^{'}}$.

\begin{figure}[h]
\begin{center}
\epsfxsize=6.5 cm \epsfysize=6.5 cm \epsfbox{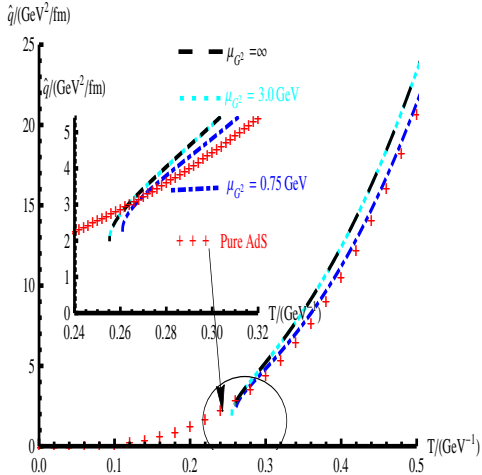}
\end{center}
\caption[]{Jet quenching parameter as a function of the temperature $T$
for $\mu_{G}=0.75 {\rm GeV}$ and $\mu_{G^2}=0.75 {\rm GeV}$, $\mu_{G^2}=3 {\rm GeV}$
and $\mu_{G^2}=\infty$ with $G_5=1.25$. The red crosses are the results of AdS-SW black hole in \cite{Liu:2006ug}. We have taken $\lambda=6\pi$ here.} \label{qhat}
\end{figure}

\begin{figure}[h]
\begin{center}
\epsfxsize=6.5 cm \epsfysize=6.5 cm \epsfbox{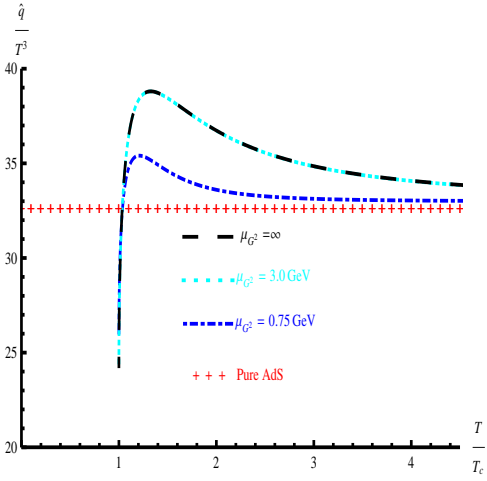}
\end{center}
\caption[]{Jet quenching parameter over cubic temperature ${\hat q}/T^3$ as a
function of $T/T_c$ for $\mu_{G}=0.75 {\rm GeV}$ and $\mu_{G^2}=0.75 {\rm GeV}$,
$\mu_{G^2}=3 {\rm GeV}$ and $\mu_{G^2}=\infty$ with $G_5=1.25$.
The red crosses are the results of
AdS-SW black hole in \cite{Liu:2006ug}. We have taken $\lambda=6\pi$ here.} \label{qhatoverT}
\end{figure}

The numerical results of the jet quenching parameter ${\hat q}$ and the ratio
of ${\hat q}/T^3$ for
$\mu_{G}=0.75 {\rm GeV}$ and $\mu_{G^2}=0.75 {\rm GeV}$, $\mu_{G^2}=3 {\rm GeV}$
and $\mu_{G^2}=\infty$ are shown in Fig.\ref{qhat} and Fig.\ref{qhatoverT}, respectively.
The results are compared with the $AdS_5$ case. For all the cases, we have taken
$\lambda=6\pi$ as in \cite{Liu:2006ug}. It is found that the jet quenching
parameter ${\hat q}$ itself does not show much differences for all the cases. It is
even hard to find much differences comparing with the $AdS_5$ case. For all these
cases, the value of ${\hat q}$ is around $5\sim 10 {\rm GeV}^2/{\rm fm}$ in
the temperature range $300 \sim 400 {\rm MeV}$, which is in agreement with the lattice result in \cite{Panero:2013pla}. However, the ratio
${\hat q}/T^3$ shows very different behavior for different cases: For the $AdS_5$ case,
the ratio is a constant, for the dynamical holographic QCD model which can describe
deconfinement phase transition, we can find that ${\hat q}/T^3$ indeed shows a peak
at the same temperature where the trace anomaly also shows a peak. For the case of $\mu_{G^2}=3 {\rm GeV}\sim \infty$, the height of the peak is around $40$ at $T=1.1 T_c$.

It is worthy of mentioning that our ${\hat q}$ is much dependent on the value of the
't Hooft coupling $\lambda$, which at the moment is a free parameter. In the recent work \cite{Burke:2013yra}, the jet quenching parameter ${\hat q}$ extracted from experiment is around $1.1 {\rm GeV}^2/{\rm fm}$ at $T=370 {\rm MeV}$ and $1.9 {\rm GeV}^2/{\rm fm}$ for $T=470 {\rm MeV}$, which is 5 times smaller than our results. This might indicate that
we should take a smaller 't Hooft coupling. However, the temperature dependent feature
is independent of the 't Hooft coupling.

\section{Jet quenching characterizing phase transition}
\label{sec-q-hat}

\begin{figure}[h]
\begin{center}
\epsfxsize=6.5 cm \epsfysize=6. cm \epsfbox{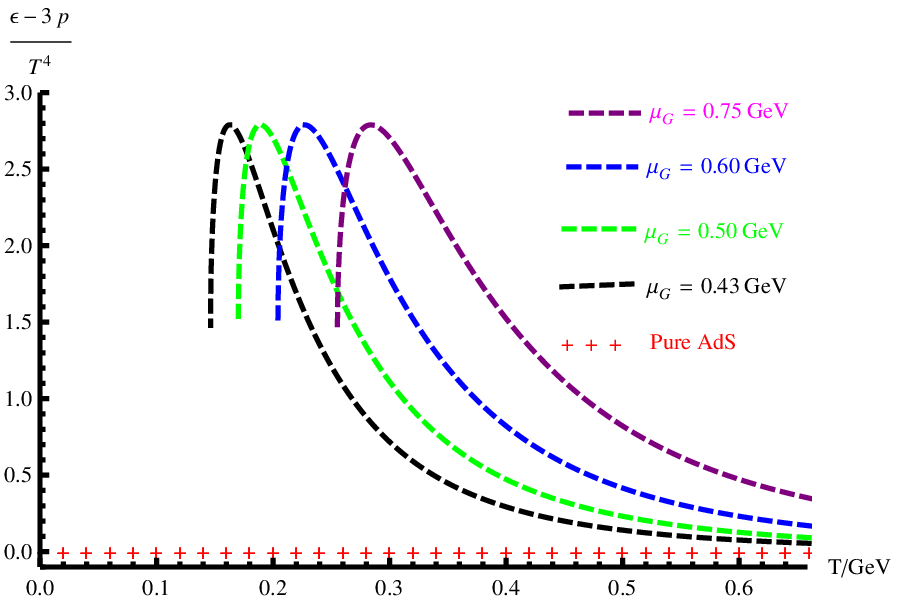} \\
\hspace*{0.1cm}
\epsfxsize=6.5 cm \epsfysize=6. cm \epsfbox{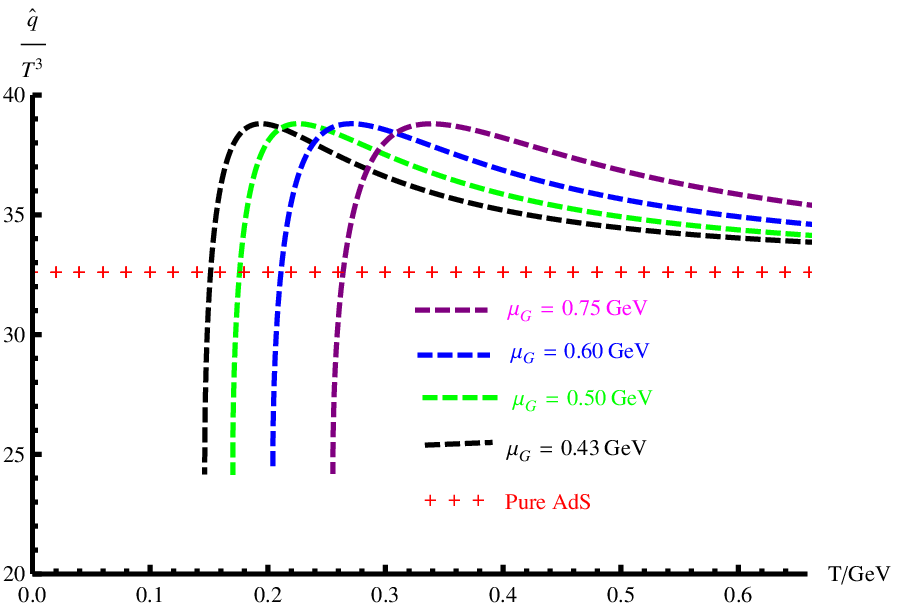} \\
\end{center}
\caption[]{$\hat{q}/T^3$ and trace anomaly $(\epsilon-3 p)/T^4$ as a function of $T$ for
different values of $\mu_G$ with $\mu_{G^2}=3 {\rm GeV}\sim \infty$. The red crosses are for $AdS_5$ case. We have taken $G_5=1.25$ and $\lambda=6\pi$ here. }  \label{T-difmuGs}
\end{figure}

\begin{figure}[h]
\begin{center}
\epsfxsize=6.5 cm \epsfysize=6. cm \epsfbox{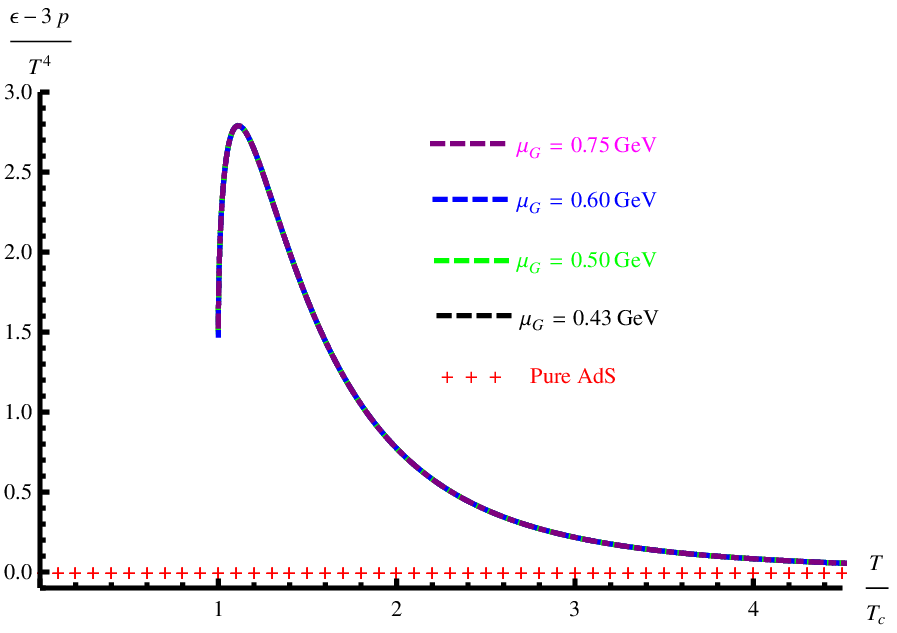} \\
\hspace*{0.1cm}
\epsfxsize=6.5 cm \epsfysize=6. cm \epsfbox{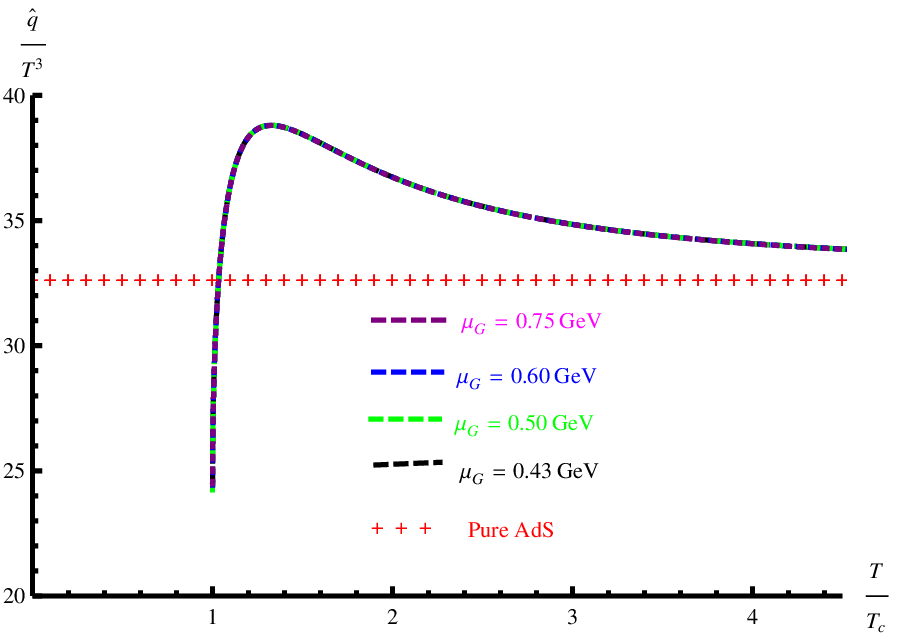} \\
\end{center}
\caption[]{$\hat{q}/T^3$ and trace anomaly $(\epsilon-3 p)/T^4$ as a function of $T/T_c$ for
different values of $\mu_G$ with $\mu_{G^2}=3 {\rm GeV}\sim \infty$. The red crosses are for $AdS_5$ case.  We have taken $G_5=1.25$ and $\lambda=6\pi$ here.}  \label{scaledTc-difmuGs}
\end{figure}

\begin{figure}[h]
\begin{center}
\epsfxsize=6.5 cm \epsfysize=6. cm \epsfbox{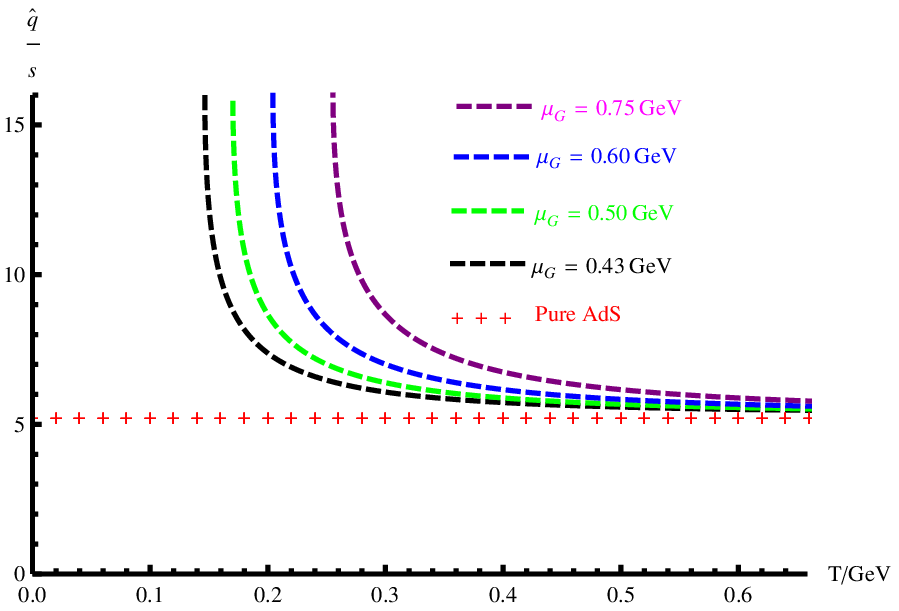} \\
\hspace*{0.1cm}
\epsfxsize=6.5 cm \epsfysize=6. cm \epsfbox{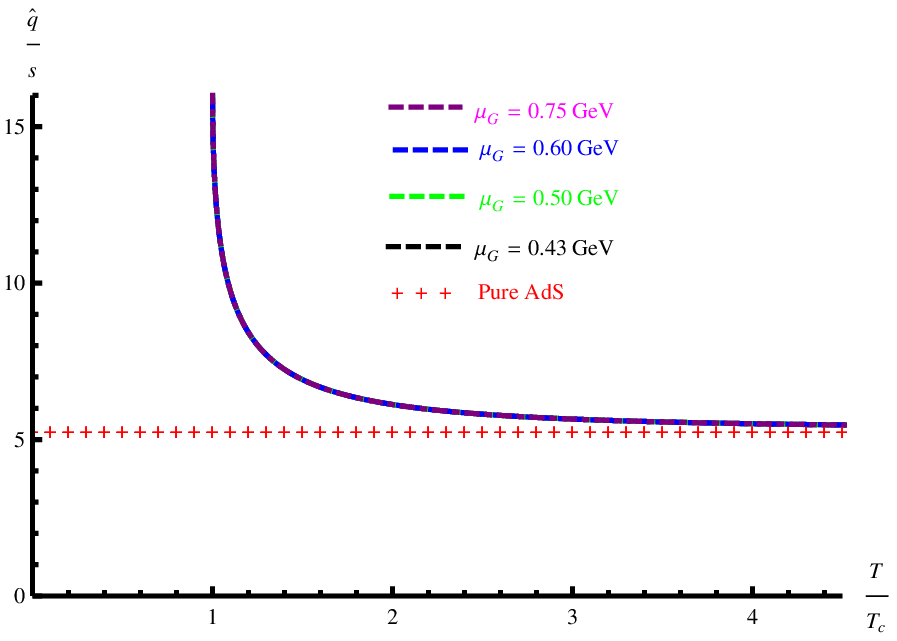} \\
\end{center}
\caption[]{$\hat{q}/s$ as a function of $T$ and $T/T_c$ for
different values of $\mu_G$ with $\mu_{G^2}=3 {\rm GeV}\sim \infty$, respectively.
The red crosses are for $AdS_5$ case.  We have taken $G_5=1.25$ and $\lambda=6\pi$ here. }  \label{qhats-difmuGs}
\end{figure}

We have observed that both $\hat{q}/T^3$ and trace anomaly $(\epsilon-3 p)/T^4$ show
a peak around the critical temperature for $\mu_G=0.75 {\rm GeV}$, which indicates
that the jet quenching parameter over cubic temperature can characterize QCD phase
transition.
In this section, we explore how different values of $\mu_G$ affect the phase transition
and jet quenching. From \cite{Li:2012ay,Li:2013oda}, $\mu_G$ is related to the linear
confinement and determines the Regge slope of the glueball spectra as well as the
string tension of the linear quark potential.

In Figs.\ref{T-difmuGs} and \ref{scaledTc-difmuGs}, we show the behavior
of $\hat{q}/T^3$ and trace anomaly $(\epsilon-3 p)/T^4$ for different values of
$\mu_G$ as a function of temperature $T$ and scaled $T/T_c$, respectively.

From Fig.\ref{T-difmuGs}, we find that for different values of $\mu_G$ (with
$\mu_{G^2}\rightarrow \infty$), the critical temperature $T_c$ increases with
$\mu_G$. We can read $T_c=146, 170, 204, 255 ~{\rm MeV}$ for $\mu_G=0.43,0.5,0.6,0.75 ~{\rm GeV}$, respectively. It is also observed that the height of the peak for
either $\hat{q}/T^3$ or $(\epsilon-3 p)/T^4$ does not change with the value of $\mu_G$,
but the width of the peak increases with $\mu_G$.

Another interesting observation from Fig.\ref{scaledTc-difmuGs} is that either $\hat{q}/T^3$ or $(\epsilon-3 p)/T^4$ as a function of scaled $T/T_c$ is not sensitive to $\mu_G$, i.e. $\hat{q}/T^3(T/T_c)$ or $(\epsilon-3 p)/T^4(T/T_c)$ overlaps for different values of $\mu_G$.

In the next section, we will investigate the nuclear modification $R_{AA}$ and elliptic
flow $v_2$, where the behavior of ${\hat q}/s$ is needed. For the case of $AdS_5$, the temperature is $T=\frac{1}{\pi z_h}$,
the entropy density takes the form of
\begin{equation}
s_{AdS_5}= \frac{1}{4 G_5}\frac{1}{z_h^3}=\frac{\pi^3}{4 G_5}T^3\simeq7.75\frac{1}{G_5}T^3,
\end{equation}
and the jet quenching parameter is given by:
\begin{equation}
{\hat q}_{AdS_5}=\frac{\pi^{3/2}\sqrt\lambda \Gamma[3/4]}{\Gamma[5/4]}T^3\simeq 7.53 \sqrt\lambda T^3.
\end{equation}
Therefore, in the $AdS_5$ limit, the ratio of jet quenching parameter over entropy density takes the value of
\begin{equation}
{\hat q}_{AdS_5}/s_{AdS_5}=0.97 G_5 \sqrt\lambda.
\end{equation}
With parameters used in our work, we have ${\hat q}_{AdS_5}/s_{AdS_5}=5.27$.
The ratio of ${\hat q}/s$ in the dynamical hQCD model as a function of $T$ and
$T/T_c$ is shown in Fig. \ref{qhats-difmuGs} compared with the $AdS_5$ result.
It is found that the ratio
of ${\hat q}/s$ reaches $AdS_5$ limit $5.27$ at high temperature, and it sharply
rises with the decreasing of $T$ and develops a peak exactly at $T_c$ with the
height $16.3$, which is about 3 times larger than its value at high temperature.
It is worthy of mentioning that the sharp rising of ${\hat q}/s$ around $T_c$
is very similar to the behavior of bulk viscosity over entropy density $\zeta/s$
as shown in \cite{LAT-xis-Meyer,LAT-xis-KT}.

Moreover, ${\hat q}/s$ as
a function of the scaled temperature $T/T_c$ overlaps for different values of $\mu_G$.

\section{Jet quenching phenomenology from holography}
\label{sec-RAA-v2}

In this section, we study the phenomenological implications of the temperature dependence for $\hat{q}(T)$ as obtained from the holography model above. The observable commonly used for jet quenching phenomenology in AA collisions is the nuclear
modification factor, $R_{AA}$, defined as the ratio between the hadron production
in $AA$ collision and that in $NN$ collision (further scaled by the expected binary
collision number). If a jet parton loses energy along its path penetrating the hot medium, one expects
a significant suppression of leading high-$p_t$ (transverse momentum) hadron production from the jet as compared with the pp collision at the same beam energy. A strong suppression was first
observed at RHIC \cite{jet-quenching} and then at LHC \cite{LHC_suppression},  with $R_{AA}$ reaching  $\sim 0.2$
in the most central collisions. Another important aspect of jet quenching is the so called geometric tomography \cite{Gyulassy:2000gk} by measuring the azimuthal angle dependence of the suppression $R_{AA}(\phi)$ where $\phi$ is the angle of the produced hadron with respect to the reaction plane. In a typical off-central collision, the hot medium on average has an almond-like geometric shape, and thus the jet in-medium path length would depend on its orientation with respect to the matter geometry, leading to nontrivial dependence of the suppression on azimuthal angle. The dominant component of the $\phi$-dependence is the second harmonic with its coefficient commonly referred to as $v_2$. Both RHIC and LHC measurements have shown a sizable $v_2$ in the high $p_t$ region where the jet energy loss should be the mechanism of generating such anisotropy~\cite{Adare:2010sp,Abelev:2012di,ATLAS:2012at,Chatrchyan:2012xq}.

The key issue we focus on here is the temperature dependence of jet-medium coupling, in particular its possible nontrivial behavior near the parton/hadron phase boundary. As was first found in \cite{Liao:2008dk}, the geometric anisotropy $v_2$ at high $p_t$ is particularly sensitive to such temperature dependence, and a simultaneous description of high $p_T$ $R_{AA}$ and $v_2$ at RHIC requires a strong enhancement of jet-medium coupling in the near-$T_c$ region. The near-$T_c$ enhancement of jet-medium interaction as a generic mechanism, has been further studied in many later works and shown to increase the jet azimuthal anisotropy with fixed overall suppression. Furthermore, how the overall opaqueness of the created fireball evolves with collisional beam energy, is also very sensitive to such temperature dependence. From RHIC to LHC, the collision beam energy increases by a little more than 10 times, and the matter density increases (in most central collisions) by a factor of about 2. Such a span from RHIC to LHC provides opportunity for determining how jet-medium interaction changes with temperature. In particularly, the near-$T_c$ enhancement predicts a visible reduction of average opaqueness of the fireball from RHIC to LHC. A number of recent analyses have consistently reported that the $R_{AA}$ data at RHIC and LHC indeed suggest about $\sim 30\%$ reduction of jet-medium interaction at LHC as compared at RHIC~\cite{Zhang:2012ie,Zhang:2012ha,Liao:2011kr,Horowitz:2011gd,Betz:2013caa,Buzzatti:2012dy,Lacey:2012kb,Zakharov:2011ws}. Therefore phenomenologically, it appears that there are now strong evidences for a nontrivial temperature dependence, in particular the near-$T_c$ enhancement, of jet-medium coupling on matter temperature.

\begin{figure}[h]
\begin{center}
\epsfxsize=6.5 cm \epsfysize=6. cm \epsfbox{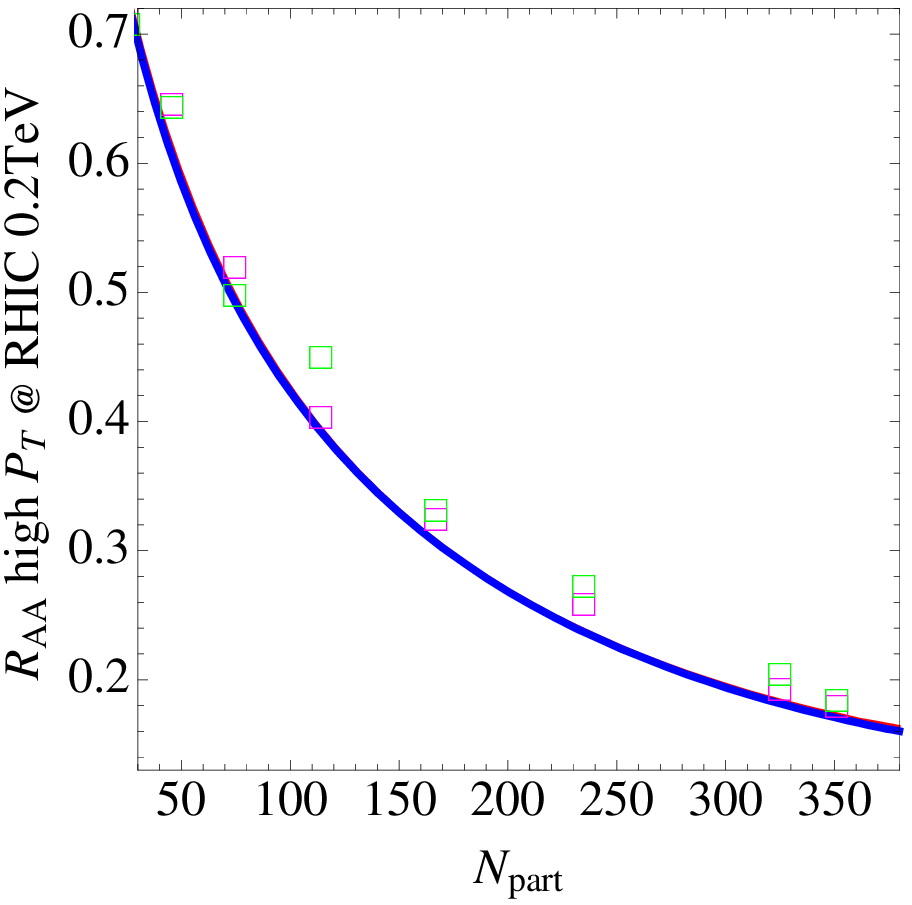} \\
\hspace*{0.1cm}
\epsfxsize=6.5 cm \epsfysize=6. cm \epsfbox{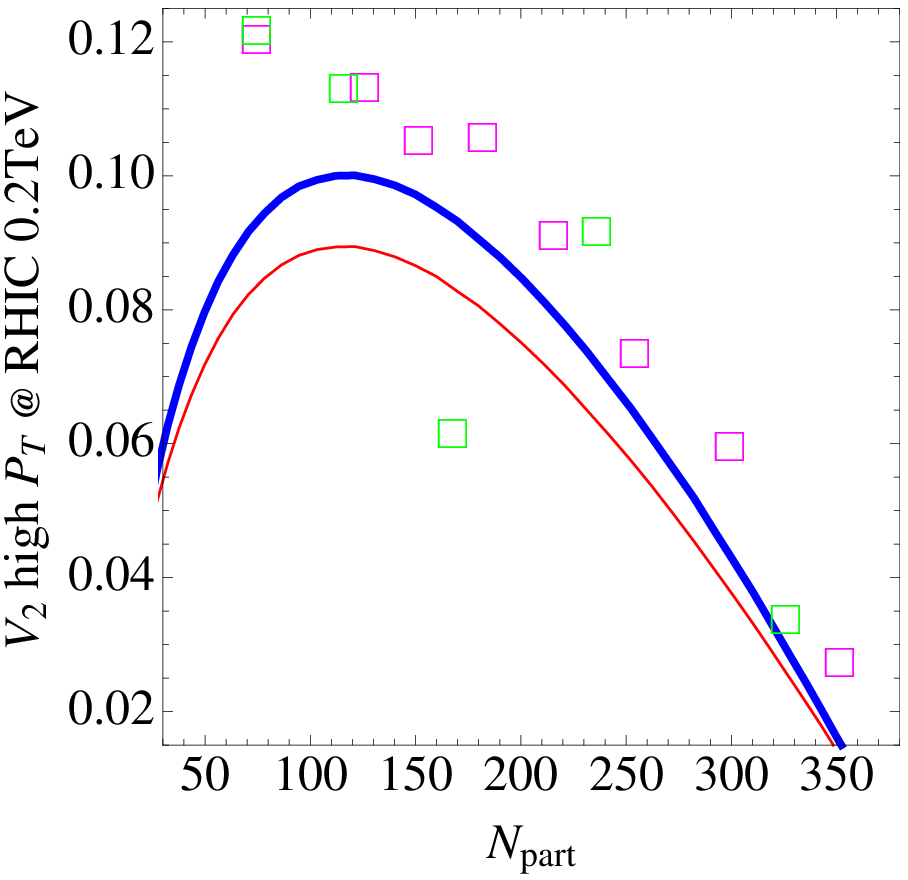} \\
\end{center}
\caption[]{The $R_{AA}$ (upper) and $v_2$ (lower) at high $p_t$ as a function of participant number $N_{part}$ for RHIC. The thick blue curves are results from non-conformal holographic model with $\hat{q}/T^3$ given in Fig.\ref{qhatoverT} ($\mu_G=0.75 {\rm GeV}, T_c=255 {\rm MeV}$) while the thin red curves are from conformal model with  $\hat{q}/T^3$ being constant. The data are PHENIX measurements of neutral pions for kinematic range in $6<p_t<9$ GeV and in $p_t >9$ GeV.  }  \label{RHIC_comparison}
\end{figure}

Theoretically, however, there has been very limited way to figure out the precise form of such T-dependence due to the highly non-perturbative nature of this temperature regime. The holographic approach provides a useful way to gain insight into this problem. In the previous Section, we've used the holographic QCD model with non-conformal dynamics  to calculate the $\hat{q}(T)$. As clearly seen Fig.\ref{qhatoverT}, the scaled jet-medium coupling $\hat{q}/T^3$ shows strong enhancement in the vicinity of $T_c$, while in contrast any conformal holographic model will show no T-dependence for the $\hat{q}/T^3$. We also emphasize that the same holographic model also describes the trace anomaly (with strong peak near $T_c$) in thermodyanics. With such  T-dependence obtained from the holographic model here,  it is of great interest to see its phenomenological implications. Here we use the simple geometric energy loss model as in \cite{Liao:2008dk,Liao:2011kr} to study the $R_{AA}$ and $v_2$ at high $p_t$ for RHIC which are most sensitive to such T-dependence. Let us assume that the final energy $E_f$ of a jet with initial energy $E_i$ after traveling an in-medium path $\vec{P}$ (specified by the jet initial spot and momentum direction) can be parameterized as $E_f = E_i \times f_{\vec P}$ with the   $f_{\vec P}$ given by
\begin{equation} \label{Eq_fP}
f_{\vec P} = \exp\left\{ - \int_{\vec P}\,  \kappa[s(l)] \, s(l) \, l dl  \right\} \ .
\end{equation}
Here $s(l)$ is the local entropy density along the jet path, while the $\kappa(s)$ represents the local jet-medium interaction strength which depends on the local density $s(l)$ (or equivalently the temperature T). We choose to explicitly separate out the density $s$ itself, and the combination $\kappa(s)\, s$   corresponds to $\hat{q}$. To implement the holographic model results for $\hat{q}$, we use $\kappa[s] = \xi \cdot  [\hat{q}/T^3]$ with $\hat{q}/T^3$ given as in Fig.\ref{qhatoverT} and with $\xi$ just one parameter to be fixed by the most central collisions $R_{AA}\approx 0.18$ for 0-5\% centrality class and then used for other computations. We use optical Glauber model to sample initial jet spots according to binary collision density and we determine a medium density from participant density with longitudinal boost-invariant expansion. It is known that there are strong initial state fluctuations, but since we are focusing on the average $R_{AA}$ and the dominant geometric anisotropy component $v_2$ which is dominantly from geometry and the current approach is reasonable (for a detailed discussions of the initial fluctuations for hard probe see e.g.~\cite{Zhang:2012ie,Zhang:2012ha,Renk:2011aa}).
In our simulations for each given impact parameter we compute the energy loss for 1 million jet paths with different initial spots and orientations, and extract the $R_{AA}$:
\begin{equation}
R_{AA} (\phi) =  <\, (f_{\vec P_{\phi}})^{n-2}  \,>_{\vec P_{\phi}} \ ,  \label{Eq_RAA}
\end{equation}
where $< \, \, >_{\vec{P}_\phi}$ means averaging over all jet paths with azimuthal orientation $\phi$ and including all sampled   initial jet production spots.  The exponent $n$ comes from reference p-p spectrum at the same collision energy: $n\approx 8.1 \ \text{and} \ 6.0 $ for $\sqrt{s}=0.2  \ \text{and} \ 2.76 $ TeV.
The so-obtained $R_{AA}(\phi)$ in each event can be further Fourier decomposed  as $R_{AA}(\phi)=R_{AA}\left[1+2 v_2 \cos(2\phi) \right] $.
The overall quenching $R_{AA}$ as well as the azimuthal anisotropy $v_2$   can then be determined.

In Fig.\ref{RHIC_comparison}, we show the results for $R_{AA}$ and $v_2$ at high $p_t$ for RHIC with the input T-dependent jet-medium interaction from our non-conformal holographic model (the thick blue curves). For comparison we also show the results from conformal model i.e. with $\hat{q}/T^3$ being constant in QGP phase (the thin red curves).  The data are from PHENIX measurements in \cite{Adare:2010sp}. As one can see, while both types of models describe the $R_{AA}$ well, the non-conformal model shows a sizable improvement over the conformal model in getting closer to the data. Of course our current non-conformal model still does not give enough anisotropy, which implies that the T-dependence of jet-medium coupling in this model may still show less near-$T_c$ enhancement than the phenomenologically favored form. Such discrepancy at quantitative level may not be unexpected due to a number of issues. After all the holographic model used here is supposed to be an effective description dual to pure gluodynamics and strictly speaking may not be suitable for direct application to full QCD phenomenology. First of all, in real QCD case with crossover transition there is the ``hadronic'' side (i.e. the sizable contribution for $\hat{q}$ when T is smaller but close to $T_c$)~\cite{Hidalgo-Duque:2013rta} which is missing in the current holographic model with 1st order transition. Furthermore, the entropy density  here (only counting the gluons essentially) is also different from full QCD where there are quarks too, and that in general would shift the peak toward lower density region in the present model. One might attempt to ``cook up'' certain extrapolative ways of accounting for such differences and thus improve the agreement with data. We however feel that would weaken the internal rigor and consistency of the holographic model approach, and would add very little to our main purpose which is not to claim success in description of data but to demonstrate the consequence of non-conformal dynamics in our holographic model on the jet energy loss phenomenology.

Let us end by reiterating our main points here: 1) there are strong non-conformal, non-perturbative dynamics going on in the near-$T_c$ region (which is modeled via holography here by introducing quadratic terms); 2) such dynamics leads to non-monotonic behavior in QGP thermodynamics as shown by the strong near-$T_c$ peak of trace anomaly (which is well modeled by holography); 3) the same dynamics leads to non-monotonic behavior in QGP transport properties and in particular strong near-$T_c$ enhancement of jet-medium coupling; 4) phenomenologically the T-dependence of $\hat{q}$ from non-conformal holographic model considerably improves the description of jet quenching azimuthal anisotropy as compared
with the conformal case.

\section{summary}
\label{sec-summary}

We have investigated QCD phase transition and jet quenching parameter ${\hat q}$
in the framework of dynamical holographic QCD model. The thermodynamical properties
in this dynamical holographic QCD model agree well with lattice results for
pure gluon system. It is found that both the
trace anomaly $(\epsilon-3 p)/T^4$ and the ratio of the jet quenching
parameter over cubic temperature ${\hat q}/T^3$ show a peak around the
critical temperature $T_c$. It is also noticed that the ratio of jet quenching
parameter over entropy density ${\hat q}/s$ sharply rises at $T_c$, which
is similar to the behavior of bulk viscosity over entropy density $\zeta/s$.
The enhancement of jet quenching parameter
around $T_c$ indicates that, like the ratio of shear viscosity over entropy
density $\eta/s$ and the ratio of bulk viscosity over entropy density $\zeta/s$,
the ratio of jet quenching parameter over entropy density ${\hat q}/s$ can
also characterize the phase transition.

The effect of jet quenching parameter enhancement around phase transition on
nuclear modification factor $R_{AA}$ and
elliptic flow $v_2$ have also been analyzed, and it is found that the
T-dependence of $\hat{q}$ from non-conformal dynamical holographic model can
considerably improve the description of jet quenching azimuthal anisotropy as
compared with the conformal case.

Here are several remarks about the dynamical holographic QCD model we used in this
work: 1) We have only considered the graviton-dilaton coupled system for the pure gluon
system, it is interesting to see in the future how the behavior of jet quenching parameter
changes by including dynamical quarks and how it will affect $R_{AA}$ and $v_2$;
2) We have only considered the gluonic matter above $T_c$, one needs to construct the thermal gas below $T_c$ in order to get the ${\hat q}/T^3$ behavior in the hadron gas;
3) One should also consider how to distinguish the energy loss of gluons and quarks
\cite{Lin:2013efa} in the framework of holography QCD.

\vskip 0.5cm
{\bf Acknowledgement}
\vskip 0.2cm

This work is supported by the NSFC under Grant No. 11275213, DFG and NSFC (CRC 110),
CAS key project KJCX2-EW-N01, K.C.Wong Education Foundation, and
Youth Innovation Promotion Association of CAS.  JL is grateful to RIKEN BNL Research Center for partial support.

\end{document}